%% file: main.tex
\documentclass[]{article}
\usepackage{spconf}
\usepackage{ amsmath, graphicx, subfig, multicol}

\title{Random Matrix Theory model for mean notch depth of the diagonally
  loaded MVDR beamformer for a single interferer case}

\twoauthors
 {Saurav R. Tuladhar, John R. Buck \sthanks{Supported by ONR grant N00014-12-1-0047}}
	{UMass Dartmouth\\
	ECE Dept.\\
	N. Dartmouth, MA}
 {Kathleen E. Wage \sthanks{Supported by ONR grant N00014-12-1-0048}}
	{George Mason University\\
	ECE Dept.\\
	Fairfax, VA}

\begin{document}
\ninept
\maketitle

\begin{abstract}
    Adaptive beamformers (ABFs) suppress interferers by placing a
    notch in the beampattern at the interferer direction. This
    suppression improves detection of a weaker signals in the presence
    of strong interferers. Hence the notch depth plays a crucial role
    in determining the adaptive gain obtained from using ABF over
    conventional beamforming. This research derives models for the
    mean notch depth of a diagonally loaded MVDR ABF for a single
    interferer case. The model describes the mean notch depth as a
    function of number of snapshots, the number of sensors in the
    array, the interferer to noise ratio (INR) level, the interferer
    direction and the diagonal loading level.  The derivation uses
    random matrix theory results on the behavior of the eigenvectors of
    sample covariance matrix. The notch depth predicted by the model
    is shown to be in close agreement with simulation results over a
    range of INRs and snapshots.
\end{abstract}

\input{newcommands}
\input{Introduction}
\input{Beamformer}
\input{Model}

\input{Results}

\input{Conclusion}
\bibliographystyle{IEEEbib}
\bibliography{IEEEabrv,myrefs}

\end{document}

%% file: newcommands.tex
॥
\newcommand{\herm}{^{\rm H}}

\newcommand{\Cov}{\boldsymbol{\Sigma}}
\newcommand{\cov}{\sigma}
\newcommand{\eval}{\gamma}
\newcommand{\Eval}{\boldsymbol{\Gamma}}
\newcommand{\evec}{\boldsymbol{\xi}}
\newcommand{\Evec}{\boldsymbol{\Xi}}

\newcommand{\sampCov}{{\bf S}}
\newcommand{\sampcov}{s}
\newcommand{\sampeval}{g}
\newcommand{\sampEval}{{\bf G}}
\newcommand{\sampevec}{{\bf e}}
\newcommand{\sampEvec}{{\bf E}}

\newcommand{\sigamp}{a}
\newcommand{\wdlmvdr}{{\bf w}_{\rm MVDR}}
\newcommand{\wt}{{\bf w}}
\newcommand{\wtscm}{\hat{\bf \wt}}
\newcommand{\dl}{\delta}
\newcommand{\wconv}{{\bf w}_{\rm c}}
\newcommand{\beampat}{\mbox{BP}}
\newcommand{\nulldepth}{\mbox{ND}}
\newcommand{\nulldepthensdl}{\ensuremath{\text{ND}_{\text{ens}_\dl}}}

\newcommand{\sv}{{\bf v}}        
\newcommand{\SV}{{\bf V}}     
\newcommand{\svlook}{\sv_0}
\newcommand{\svint}{\sv_1}
\newcommand{\tsp}{^{\sv T}}
\newcommand{\inv}{^{-1}}
\newcommand{\limrmt}{\lim_{RMT}}
\newcommand{\eig}{\operatorname{eig}}
\newcommand{\diag}{\operatorname{diag}}

\newcommand{\red}[1]{\textcolor{red}{#1}}
\newcommand{\blue}[1]{\textcolor{blue}{#1}}
\newcommand{\green}[1]{\textcolor{green}{#1}}

\newcommand{\sect}{Sec.}
\newcommand{\fig}{Fig.}
\newcommand{\eqn}{Eq.}

%% file: Introduction.tex
\section{Introduction}
\label{sec:intro}
A common array processing problem is to detect a low power source in
presence of high power interferers. A conventional beamformer (CBF)
produces a static beampattern which attenuates interferers by a fixed
amount at each bearing. At the output, the weaker signal of interest
will be masked by the higher power interferer and undermines
detection. Alternatively, adaptive beamformers (ABF) can suppress
interferers by placing deep notches in the beampattern in the
interferer direction. ABFs rely on the knowledge of the data
covariance matrix to compute the beamformer weights. In reality, the
ensemble covariance matrix (ECM) for data is unknown \emph{a
  priori}. The traditional approach is to replace the ECM by the
sample covariance matrix (SCM) to compute the beamformer weights.

A class of sample matrix inversion (SMI) ABFs involve inverting the
SCM to compute the beamformer weights
\cite[Sec. 7.3]{vtree2002oap}. If the number of snapshots ($L$) is
less than or approximately equal to the number of array sensors ($N$),
the SCM is unstable or ill-conditioned for inversion. A common
approach is diagonally loading the SCM to make it invertible for
computing the ABF weights. The minimum variance distortionless
response (MVDR) beamformer is one of the most extensively used SMI
ABFs \cite{capon1969mvdr}. The main focus of this paper is to
characterize the mean notch depth of a diagonally loaded (DL) MVDR
ABF.

Prior work by Richmond \cite{richmond2000mvdr} derived expressions for
the mean and the variance of the SCM based MVDR beampattern. However
the derivation in \cite{richmond2000mvdr} only considers the snapshot
sufficient case ($N < L$) and does not include diagonal loading. More
recently, Buck and Wage \cite{buck2012dmr} used Random Matrix Theory
(RMT) results to develop a model for the mean notch depth of the
dominant mode rejection (DMR) ABF. DMR is a variant of the MVDR ABF
that uses a constrained SCM instead of diagonal loading
\cite{abraham1990beamforming}. Mestre and Lagunas
\cite{mestre2006finite} have used RMT results to derive a
deterministic expression for asymptotic output
signal-to-interferer-plus-noise ratio (SINR) of a DL-MVDR ABF. Their
analysis is focused on deriving an estimator for the optimum loading
factor ($\dl$). Similarly, Pajovic et al. \cite{pajovic2012mpdr} used
RMT results to derive an analytic expression for the output power of a
DL minimum power distortionless response (MPDR) beamformer. MPDR
assumes source signal is present in the training data
\cite[Sec. 6.2.4]{vtree2002oap}.

The results presented in this paper are similar in spirit to the work
in \cite{buck2012dmr}, but for the DL-MVDR ABF also considered in
\cite{mestre2006finite}. Recent results from RMT are used to derive an
approximate model for the mean notch depth of a the DL-MVDR. The model
will describe the notch depth as a function of the diagonal loading
level ($\dl$) in addition to the number of snapshots ($L$), number of
sensors ($N$), the interferer to noise ratio (INR), and the interferer
location ($\theta_1$).

The rest of the paper is organized as follows: The next section
describes the MVDR beamformer and defines related
terminologies. \sect{}~\ref{sec:model} summarizes the notch depth model
derivation. The simulation results are discussed in
\sect{}~\ref{sec:results}, followed by a brief conclusion in \sect{}~\ref{sec:conclusion}


%% file: Beamformer.tex
\section{The MVDR Beamformer}
\label{sec:mvdr_abf}
The MVDR beamformer is one of the most extensively used ABFs
\cite{vtree2002oap,capon1969mvdr}. 
The weight vector for the MVDR ABF steered to bearing direction $\theta_0$ is,
\begin{equation}
    \label{eq:mvdr_wt}
    \wt  =  \Cov\inv \svlook / \left(\svlook\herm \Cov\inv \svlook\right)
\end{equation}
where $\Cov$ is the $N\times N$ ECM and $\svlook = \sv(\theta_0)$ is the
array steering vector corresponding to the look direction $\theta_0$. Assuming a stationary narrowband interferer with power $\cov_1^2$ at
bearing $\theta_1$ and unit power white background noise, the ECM is
\begin{equation}
    \label{eq:ecm}
    \Cov = \cov_1^2\svint\svint\herm + {\bf I} = \sum_{i=1}^{N}\eval_i\evec_i\evec_i\herm,
\end{equation}
where $\eval_1 > \eval_2 = \ldots \eval_N = 1$ are the eigenvalues and
$\evec_i$ are the corresponding eigenvectors. In the
 single interferer case $\evec_1 = \svint / \sqrt{N}$, i.e., the principal eigenvector
is a scaled version of the interferer steering vector ($\svint$). The
MVDR ABF places a
notch in the direction corresponding to $\evec_1$.

In practice the ECM is estimated by computing the SCM ($\sampCov$)
from $L$ data snapshot vectors ${\bf x}_l$,
\[
 \sampCov = \frac{1}{L}\sum_{l=1}^{L}{\bf x}_l{\bf x}_l\herm = \sum_{i=1}^{N}\sampeval_i\sampevec_i\sampevec_i\herm.
\]
where ${\bf x}_l = \sigamp_{l}\svint + {\bf n}_l$ such that $
\sigamp_{l} \sim {\cal CN}(0,\cov_1^2)$ and ${\bf n}_l \sim {\cal
  CN}(0, {\bf I})$. Since the noise power is unity, the INR is equal to
$\cov_1^2$. Similarly $\sampeval_1 > \sampeval_2 \geq \ldots \geq
\sampeval_N$ are the eigenvalues and $\sampevec_i$ are the
eigenvectors of the SCM. The DL SCM is computed as $\sampCov_\dl =
\sampCov + \dl{\bf I}$ where $\dl > 0$. The eigenvectors are invariant
to DL. Hence the sample principal eigenvector $\sampevec_1$ estimates the
interferer direction. The DL-MVDR ABF weights ($\wtscm_\dl$) are
computed by replacing $\Cov$ with  $\sampCov_\dl$ in \eqref{eq:mvdr_wt}.

\subsection{Notch Depth}
\label{sec:notch-depth}
The notch depth is defined as the magnitude of the beampattern at true
interferer direction, i.e., $\nulldepth = |\wt \herm \svint|^2$. The
ensemble notch depth for the DL-MVDR is
\begin{align}        
    \label{eq:ens_nd}
    \nulldepth_{\text{ens}_\dl} = \frac{\cos^2(\svlook,\svint)(1+\dl)^2}{\left |(1 + \dl
          + N\cov_1^2\sin^2(\svlook,\svint)\right |^2},
\end{align}
where $\cos^2(\svlook,\svint)$ is the generalized cosine between
$\svlook$ and $\svint$ as defined in
\cite{cox1973resolving}. $\nulldepth_{\text{ens}_\dl}$ is the ideally
achievable notch depth assuming the ECM is known. Computing the
weights with the SCM results in a notch depth
${\nulldepth}_\dl$ which is shallower than the ensemble $\nulldepth_{\text{ens}_\dl}$. The mismatch between the sample and
ensemble principal eigenvectors is the main cause of this loss in
notch depth \cite[Sec. 5]{wage2013dmr}. RMT has results on bias of
eigenvectors ($\sampevec_i$) of the SCM, which will be used to derive
the mean notch depth model in the next section.


%% file: Model.tex
\section{Model}
\label{sec:model}
This section derives two models for the DL-MVDR ABF notch depth.  The
models characterize notch depth as a function of the number of sensors $N$,
the number of snapshots $L$, the INR ($\cov_1^2$), the interferer direction
($\theta_1$) and the diagonal loading level ($\dl$). The first model
treats snapshots ($L$) as the independent variable and
INR ($\cov_1^2$) as a parameter. The second model characterizes the ND
as a function of the INR ($\cov_1^2$) while treating the snapshots ($L$)
as a parameter. The derivation uses the RMT results on the fidelity of
the sample principal eigenvectors.

The first part of the RMT result on the eigenvectors of SCM gives an
expression for the magnitude of the projection between the sample and
the ensemble principal eigenvector,
\cite{paul2007asymptotics}\cite{benaych2011eigen}\cite{johnstone2009consistency},
\begin{equation}
    \label{eq:rmt_result}
    |\sampevec_1\herm\evec_1|^2 \overset{a.s.}{\to}
    \begin{cases}
        \frac{1 - c/(N\sigma_1^2)^2}{1 + c/(N\sigma_1^2)} &  \cov_1^2 > \sqrt{c}/N \\
        0       &  \cov_1^2 \leq \sqrt{c}/N 
    \end{cases}
\end{equation}
where $c = N/L$. This result holds in the RMT asymptotic sense, i.e., $N,
L \rightarrow \infty, N/L \rightarrow c$. It implies that for a
sufficiently strong interferer ($N\cov_1^2 > \sqrt{c}$) the sample
principal eigenvector $\sampevec_1$ is a biased estimate of its
ensemble counterpart.

The second part of the result states that the noise eigenvectors are
uniformly distributed over a unit sphere
\cite[Thm. 6]{paul2007asymptotics}. This implies the magnitude of
projection of sample principal eigenvector on the orthogonal vector
$\evec_\perp$ is,
\begin{equation}
\label{eq:rmt_noise_result}
|\sampevec_1\herm\evec_\perp|^2 = (1 - |\sampevec_1\herm\evec_1|^2)/(N-1).
\end{equation}

Further, the look direction steering vector $\svlook$ can be decomposed into two
orthogonal unit vectors $\evec_1$ and $\evec_\perp$,
\begin{equation}
    \label{eq:sv_orth}
    \svlook = \alpha \evec_1 + \beta \evec_\perp     
\end{equation}
where $\alpha = \sqrt{N}\cos(\svlook,\svint)$ and $\beta =
\sqrt{N}\sin(\svlook,\svint)$.

\subsection{Notch Depth vs Snapshots}
\label{sec:notch-depth-vs-snapshots}
The derivation of notch depth vs snapshots model begins from the
expression for ${\nulldepth}_\dl$ by substituting for $\svlook$ from
\eqref{eq:sv_orth} and setting $\svint = \sqrt{N}\evec_1$. This
substitution results in expressions containing quadratic terms
$|\sampevec_1\herm\evec_1|^2$ and
$|\sampevec_1\herm\evec_\perp|^2$. The two terms are then replaced
using RMT results in \eqn~\eqref{eq:rmt_result}~and
\eqref{eq:rmt_noise_result}.  Collecting common terms in $L$ and
factoring appropriately simplifies the notch depth expression to
\begin{equation}
\label{eq:nd_L_model}
\nulldepth_{\dl}
\approx \cos^2(\svlook,\svint)(1 + \dl)^2\frac{|f_3(L)f_2(L)|^2}{|f_1(L)|^2},
\end{equation}
where
\begin{align}
\label{eq:nd_L_factor}
    \begin{split}
        f_1(L) &=  N + L(1 + \dl + (N\cov_1^2)\sin^2(\svlook,\svint))\\
        f_2(L) &= \sqrt{L} - \sqrt{N}\cov_1^{-1}\cot(\svlook,\svint)   \\
        f_3(L) &= \sqrt{L} -\frac{\sqrt{N}\cov_1}{1 + \dl}\tan(\svlook,\svint).
    \end{split}
\end{align}
This derivation assumes that the array is sufficiently long ($N \gg
1$), the interferer power is strong enough ($N\cov_1^2 \gg 1$) and the
interferer lies outside the main lobe of CBF
($\cov_1^2\tan^2(\svlook,\svint)\gg 1$). For the snapshot sufficient case
of $ c \leq 1$ DL is constrained to $\dl > (1 - \sqrt{c})^2 $
\begin{figure}[t]
    \centering
    \centerline{\includegraphics[scale=0.5]{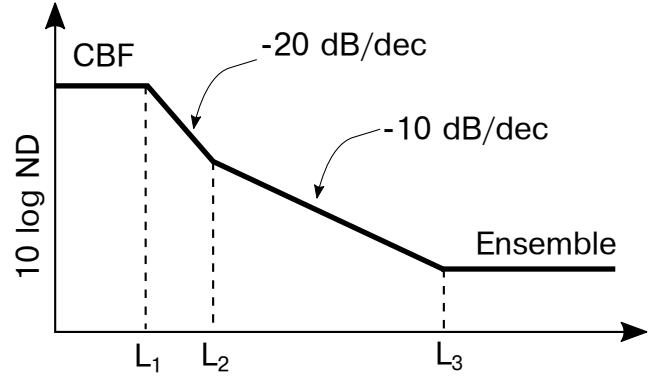}}
    \caption{Notch depth vs snapshots model}
    \label{fig:nd_vs_L}
\end{figure}
The model in \eqn{}~\eqref{eq:nd_L_model} can be visualized as a
linear piecewise function of $L$ in a log-log scale. This
interpretation of the model is similar to Bode plot approach to
interpret system transfer functions \cite{valkenburg1958network}. The
same approach was used to interpret DMR notch depth model in
\cite[Sec. 3.1]{buck2012dmr}. As $L$ increases each factor in
\eqref{eq:nd_L_factor} becomes significant over a different range of
values of $L$. The increase in magnitude of each factor dictates the
slope of the linear piecewise function. The values of $L$ for which
the summands in each factor become equal predict the breakpoints,
\begin{align*}
    L_1 & = N/\left(\dl + \cov_1^2 N\sin^2(\svlook,\svint)\right)
    & \mbox{(2nd order)}\\
    L_2 & = N \cot^2(\svlook,\svint)/\cov_1^2   & \mbox{(1st order)}\\
    L_3 & = N \cov_1^2 \tan^2(\svlook,\svint)/(1 + \dl)^2  & \mbox{(1st order)}.
\end{align*}
The resulting model is shown in \fig{}~\ref{fig:nd_vs_L}. The model
predicts that for smaller values of $L$, the DL-MVDR notch depth
reduces to the CBF case. Gathering more snapshots ($L > L_1$) results
in increased nulling. With a sufficiently large number of snapshots,
notch depth converges to the ensemble value. The breakpoint values
$L_3$ suggests that increasing the diagonal loading ($\dl$) reduces
the snapshots required to achieve the DL ensemble notch depth
($\nulldepthensdl$).

\subsection{Notch Depth vs INR}
\label{sec:notch-depth-vs-inr}
\begin{figure}[t]
    \centering
    \centerline{\includegraphics[scale=0.48]{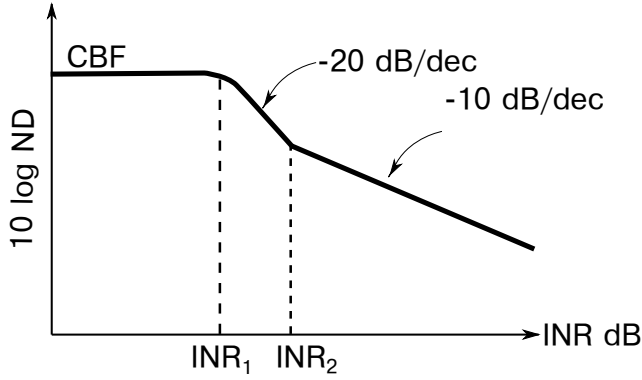}}
    \caption{Notch depth vs INR model}
    \label{fig:nd_vs_INR}
\end{figure}

The notch depth vs INR model is developed following similar steps used
in \sect{}~\ref{sec:notch-depth-vs-snapshots}. This model collects
the terms common in $\cov_1^2$ and factors appropriately to obtain
\begin{equation}
\label{eq:nd_vs_inr}
 \nulldepth_{\dl} \approx 
  \cos^2(\svlook,\svint) \frac{\left|\cov_1 \sqrt{c} \tan(\svlook,\svint) - (1 + c + \dl)\right|^2}{\left| N \cov_1^2
    \sin^2(\svlook,\svint) + (1 + c + \dl) \right|^2}
\end{equation}
The notch depth model in is once again interpreted using the same
approached discussed in
\sect{}~\ref{sec:notch-depth-vs-snapshots}. This approach models
the notch depth as a piecewise linear function of INR
($\cov_1^2$) in a log-log scale as shown in
\fig{}~\ref{fig:nd_vs_INR}.
The two dyadic factors in \eqn{}~\eqref{eq:nd_vs_inr} predict the
breakpoint values of INR to be,
\begin{align*}
   \text{INR}_1 & = (1 + c +\dl)/N\sin^2(\svlook,\svint)\\
   \text{INR}_2 & = (1 + c +\dl)^2/c \tan^2(\svlook,\svint).
\end{align*}

%% file: Results.tex
\section{Simulation Results and Discussions}
\label{sec:results}
This section compares the estimated notch depth from computer
simulations of DL-MVDR and the model predictions. The
simulations were performed for a uniform linear array with $N = 50$
sensors. A single stationary interferer was assumed to be at bearing
$u_1 = \cos(\theta_1) = 0.06$, which is the location of the
peak of the CBF first sidelobe.

\begin{figure}[!tbh]
    \centering 
    \includegraphics[scale=0.31]{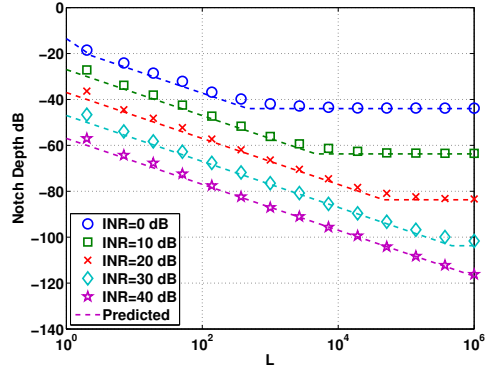}   
     \caption{Notch depth vs snapshots simulation results compared to
      model prediction for $\dl = 0.5$.}
    \label{fig:ND_vs_L}
\end{figure}

\fig{}~\ref{fig:ND_vs_L} compares the notch depth as a function of
snapshots $L$, predicted by the RMT model in
\eqn{}~\eqref{eq:nd_L_model} with the notch depth estimated from
simulation for different INR levels. The dashed lines represent the
notch depth predicted by the model. The discrete markers represent the
average notch depth obtained from a 500 trial Monte Carlo
experiment. The black markers represent the ensemble notch depth
($\nulldepthensdl$) at each INR ($\cov_1^2$) level. The model
predicted notch depth matches the averaged notch depth observed in the
simulations.

\begin{figure}[!tphb]
    \centering
    \includegraphics[scale=0.31]{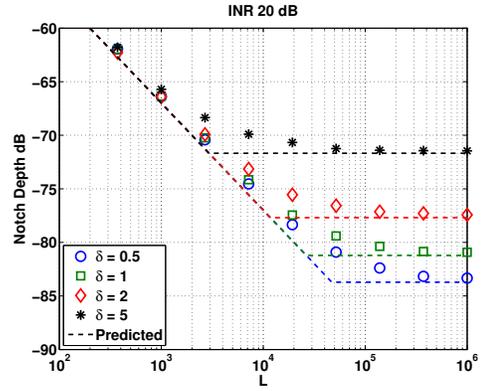}
    \caption{Notch depth vs snapshots for INR of 20 dB at different
      diagonal loading levels}
    \label{fig:ND_vs_L_delta}
\end{figure}

For most practical array sizes and strong interferers, the breakpoints
predicted in \eqref{eq:nd_L_factor} are such that $L_1 < 1$, $L_2
\approx 1$ and $L_3$ is practically unattainable. Hence, the first two
breakpoints are limited to a theoretical interpretation of the
model and are not observed in \fig{}~\ref{fig:ND_vs_L}. In practice the
region of operation lies between $L_2$ and $L_3$ where notch depth
grows by 10 dB for every decade increase in $L$.

\fig{}~\ref{fig:ND_vs_L_delta} compares the notch depth as a function
of snapshots at INR of 20 dB for different DL levels.. The figure
indicates that higher DL allows the DL-MVDR ABF to approach
ensemble ND with fewer snapshots as predicted by expression for
$L_3$. However increasing the loading level also makes the ensemble notch
depth $\nulldepthensdl$ shallower.

\begin{figure}[!tbh]
    \centering 
    \includegraphics[scale=0.31]{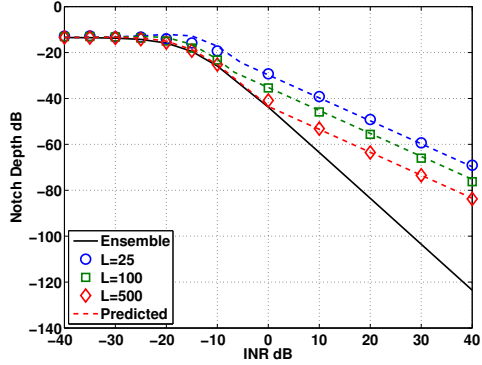}
    \caption{Notch depth vs INR simulation results compared to model
      prediction for $\dl = 0.5$}
    \label{fig:ND_vs_INR}
\end{figure}

\begin{figure}[!tbh]
    \centering
    \includegraphics[scale=0.31]{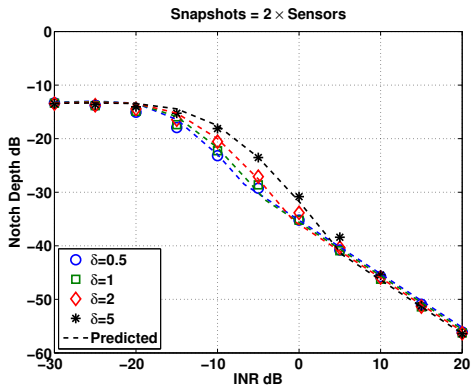}
    \caption{Notch depth vs INR for L = 2N at different diagonal
      loading levels}
    \label{fig:ND_vs_INR_delta}
\end{figure}

\fig{}~\ref{fig:ND_vs_INR} compares the notch depth as a
function of INR, predicted by the RMT model in
\eqn{}~\eqref{eq:nd_vs_inr} and the notch depth estimated from
simulations for different snapshots $L$. The dashed lines represent
notch depth predicted by the model. The discrete markers represent the
average notch depth obtained from 500 trial Monte Carlo
experiments. The solid line represents the ensemble behavior over the
range of INR. Again, the model predicted notch depth matches the
averaged notch depth observed in the simulations.

The range of INRs between $\text{INR}_1$ and $\text{INR}_2$ is where
the DL-MVDR ABF is adapting to the change in interferer power. The notch depth
grows by 20 dB for every 10 dB rise in interferer power once the
interferer power is higher than $\text{INR}_1$. The
interferer is suppressed more, the stronger it becomes. Once the
interferer power exceeds $\text{INR}_2$, the notch depth growth merely
keeps up with interferer power rise. Consequently the interferer power
in the beamformer output remains unchanged. Hence there is no
additional adaptive gain in using DL-MVDR ABF in this range of INR
levels.

\fig{}~\ref{fig:ND_vs_INR_delta} compares the notch depth as function
of INR for snapshots $L = 2N$ among different DL levels. The figure
shows that the DL-MVDR ABF with higher DL can suppress interferers
with higher power $\cov_1^2$. On the other hand, increasing DL means
that the INR must grow larger before the DL-MVDR ABF actually begins
adapting. $\text{INR}_1$ is effectively the minimum interferer power
required for the DL-MVDR ABF to depart from CBF performance and begin
adapting.


%% file: Conclusion.tex
\section{Conclusion}
\label{sec:conclusion}
This paper presents RMT based models for the mean notch depth of a
DL-MVDR ABF in a single interferer case. The simulation results verify the
accuracy of the notch depth predicted by the two models. The derived
models indicate that increasing the diagonal loading reduces
snapshots required to converge to the ensemble notch depth. Similarly,
the ability to suppress an interferer with higher power increases with higher
diagonal loading. The improved suppression comes at a cost of
shallower DL ensemble notch depth.
